# MULTIPHONON GIANT RESONANCES

C.A. BERTULANI
*Instituto de Física, Universidade Federal do Rio de Janeiro*
*21945-970, Rio de Janeiro, RJ, Brazil*
*E-mail: bertu@if.ufrj.br*

A new class of giant resonances in nuclei is discussed, i.e., giant resonances built on other giant resonances. These resonances are observed with very large cross sections in relativistic heavy ion collisions. A great experimental and theoretical effort is underway to understand the reaction mechanism which leads to the excitation of these states in nuclei, as well as the better microscopic understanding of their properties, e.g., strength, energy centroids, widths, and anharmonicities.

## 1 Giant Resonances

### *1.1 Single giant resonances*

Giant resonances in nuclei were first observed in 1937 by Bothe and Gentner [1]who observed an unexpectedly large absorption of 17.6 MeV photons (from the $^{7}Li(p,\ \gamma)$ reaction) in some targets. These observations were later confirmed by Baldwin and Klaiber (1947) with photons from a betatron. In 1948 Golhaber and Teller [2] interpreted these resonances (named by isovector giant dipole resonances) with a hydrodynamical model in which rigid proton and neutron fluids vibrate against each other, the restoring force resulting from the surface energy. Steinwendel and Jansen [3] later developed the model, considering compressible neutron and proton fluids vibrating in opposite phase in a common fixed sphere, the restoring force resulting from the volume symmetry energy. The standard microscopic basis for the description of giant resonances is the Random Phase Approximation (RPA) in which giant resonances appear as coherent superpositions of one-particle one-hole ($1p1h$) excitations in closed shell nuclei or two quasi-particle excitations in open shell nuclei (for a review of these techniques, see, e.g., ref. [4]).

The isoscalar quadrupole resonances were discovered in inelastic electron scattering by Pitthan and Walcher (1971) and in proton scattering by Lewis and Bertrand (1972). Giant monopole resonances were found later and their properties are closely related to the compression modulus of nuclear matter. Following these, other resonances of higher multipolarities and giant magnetic resonances were investigated. Typical probes for giant resonance studies are (a) $\gamma$'s and electrons for the excitation of GDR (isovector giant dipole resonance), (b) $\alpha$-particles and electrons for the excitation of isoscalar GMR (giant monopole resonance) and GQR (giant quadrupole resonance), and (c) $(p, n)$, or ($^3$He, t), for Gamow-Teller resonances, respectively.

*1.2 Multiphonon resonances*

Inelastic scattering studies with heavy ion beams have opened new possibilities in the field (for a review the experimental developments, see ref. [5]). A striking feature was observed when either the beam energy was increased, or heavier projectiles were used, or both. This is displayed in figure 1, where the excitation of the GDR in $^{208}Pb$ was observed in the inelastic scattering of $^{17}O$ at 22 MeV/nucleon and 84 MeV/nucleon, respectively, and $^{36}Ar$ at 95 MeV/nucleon. What one clearly sees is that the "bump" corresponding to the GDR at 13.5 MeV is appreciably enhanced. This feature is solely due to one agent: the electromagnetic interaction between the nuclei. This interaction is more effective at higher energies, and for increasing charge of the projectile.

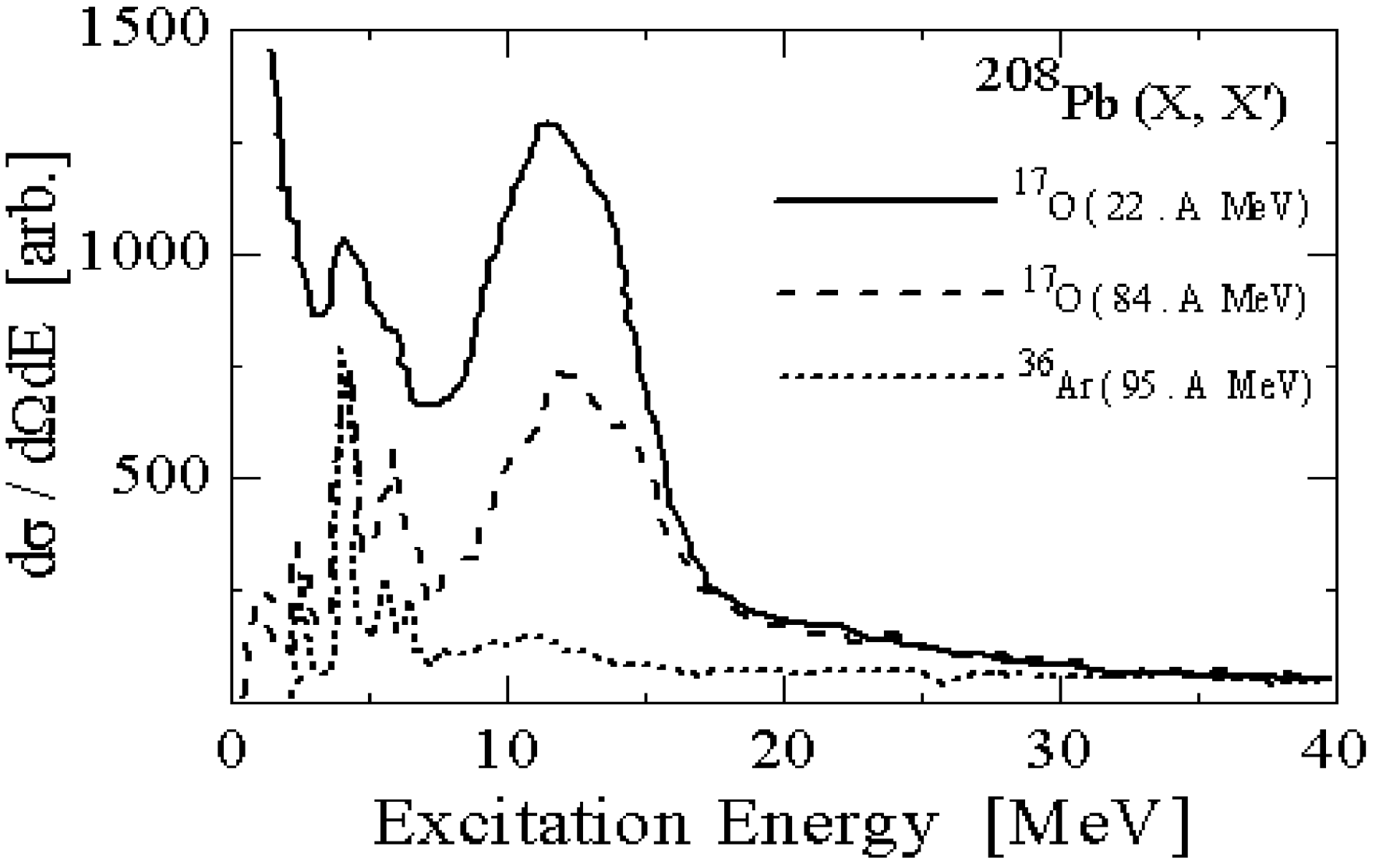

*Figure 1*

In ref. [6] it was noted that the excitation probabilities of the GDR in heavy ion collisions approach unity at grazing impact parameters. It was further shown that, if double GDR resonance (i.e. a GDR excited on a GDR state) exists then the cross sections for their excitation in heavy ion collisions at rel-ativistic energies are of order of hundreds of millibarns. This calculation was based on the semiclassical approach, appropriate for heavy ion scattering at high incident energies, and the harmonic oscillator model for the giant resonances. The semiclassical model treats the relative motion between the nuclei classically while quantum mechanics is used for the internal degrees of freedom.

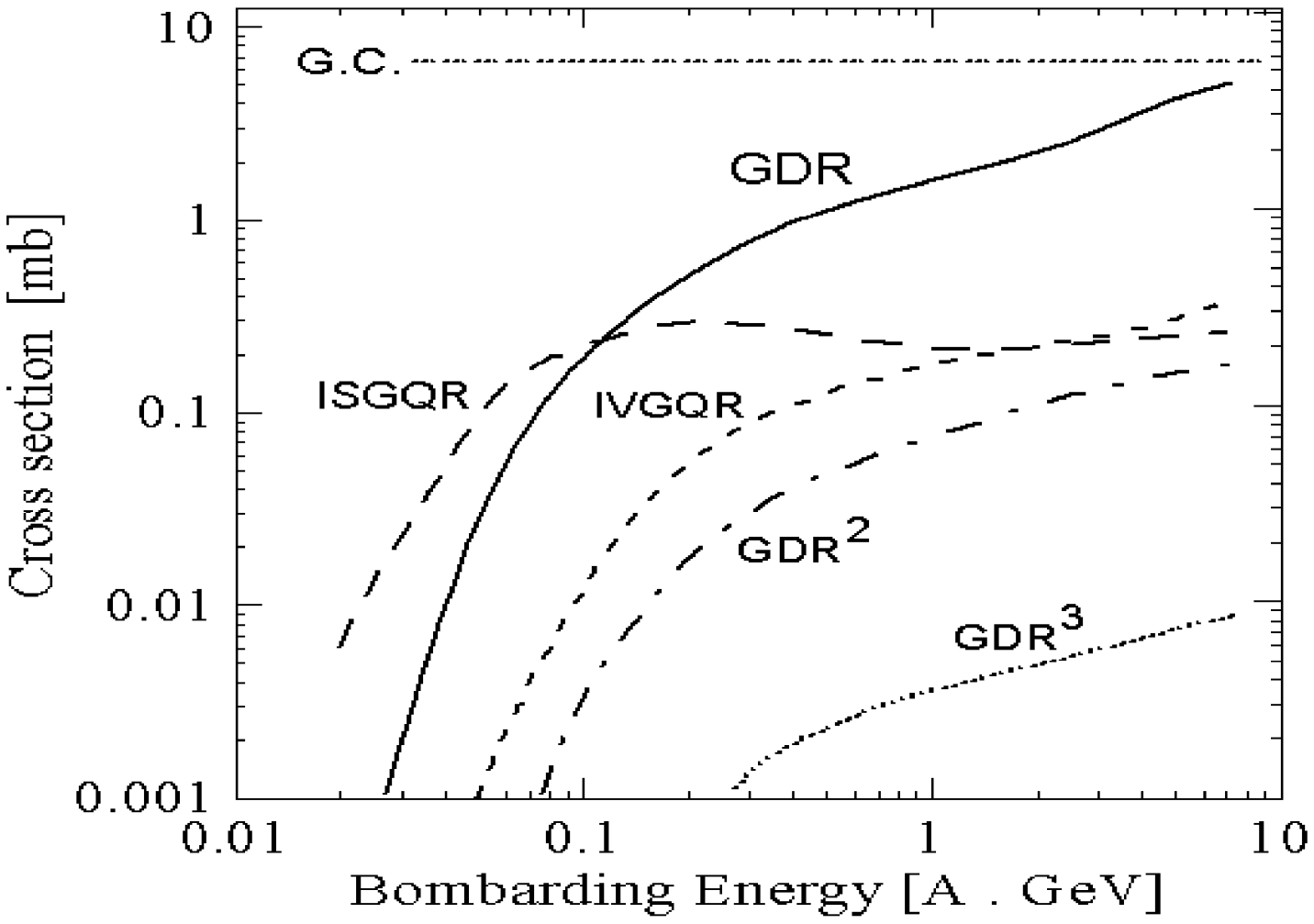

*Figure 2*

In the harmonic picture for the internal degrees of freedom the GDR is the first excited state in a harmonic well, the GDR$^2$, or DGDR (double GDR), is the second state, and so on. In ref. [7] it was shown that the excitation probabilities and cross sections are directly proportional to the photonuclear cross sections for a given electric (E) and magnetic (M) multipolarity. For an impact parameter $b$, excitation energy $E$, and a multipolarity $\pi\lambda$ ($\pi$ = E or M) the excitation probabilities are given by

$$P_{\pi\lambda}(E,\ b) = \frac{1}{E}\ N_{\pi\lambda}(E,\ b)\ \sigma_\gamma^{\pi\lambda}(E) \tag{1}$$

where $\sigma_\gamma^{\pi\lambda}(E)$ is the photonuclear cross sections for the photon $E$ and multipolarity $\pi\lambda$. The total photonuclear cross section is $\sigma_\gamma(E) = \sum_{\pi\lambda} \sigma_\gamma^{\pi\lambda}(E)$. In the semiclassical approach, the "equivalent photon numbers" $N_{\pi\lambda}(E,\ b)$ are given analytically[7]. A quantum mechanical derivation of the excitation amplitudes in relativistic Coulomb excitation shows that eq. (1) can also be obtained by using the saddle point approximation in the DWBA integrals [8]. The total Coulomb excitation cross sections can be obtained by an integration of eq. (1) over the impact parameter $b$, including a factor, $T(b)$, which accounts for the strong absorption at small impact parameters: $\sigma_{\pi\lambda}(E) = 2\pi \int db\ b\ T(b) P_{\pi\lambda}(E,\ b)$. The impact parameter integral can also be performed analytically and the equivalent photon numbers $n_{\pi\lambda}(E) = 2\pi \int db\ b\ N_{\pi\lambda}(E,\ b)$ are given in refs [7,8].

The cross section for the excitation of a giant resonances is obtained from these expressions, by using the experimental photonuclear absorption cross section for $\sigma_\gamma^{\pi\lambda}(E)$ in eq. (1). One problem with this procedure is that the experimental photonuclear cross section includes all multipolarities with the same weight: $\sigma_\gamma^{\rm exp}(E) = \sum_{\pi\lambda} \sigma_\gamma^{\pi\lambda}(E)$, while the calculation based on eq. (1) needs the isolation of $\sigma_\gamma^{\pi\lambda}(E)$. This can be done only marginally, except in some exclusive measurements. Generally, one finds in the literature the $(\gamma,\ n)$, $(\gamma,\ 2n)$, and $(\gamma,\ 3n)$ cross sections, which include the contribution of all multipolarities in the giant resonance energy region. A separation of the different multipolarities can be obtained roughly by use of sum rules, or some theoretical model for the nuclear response to a photo excitation.

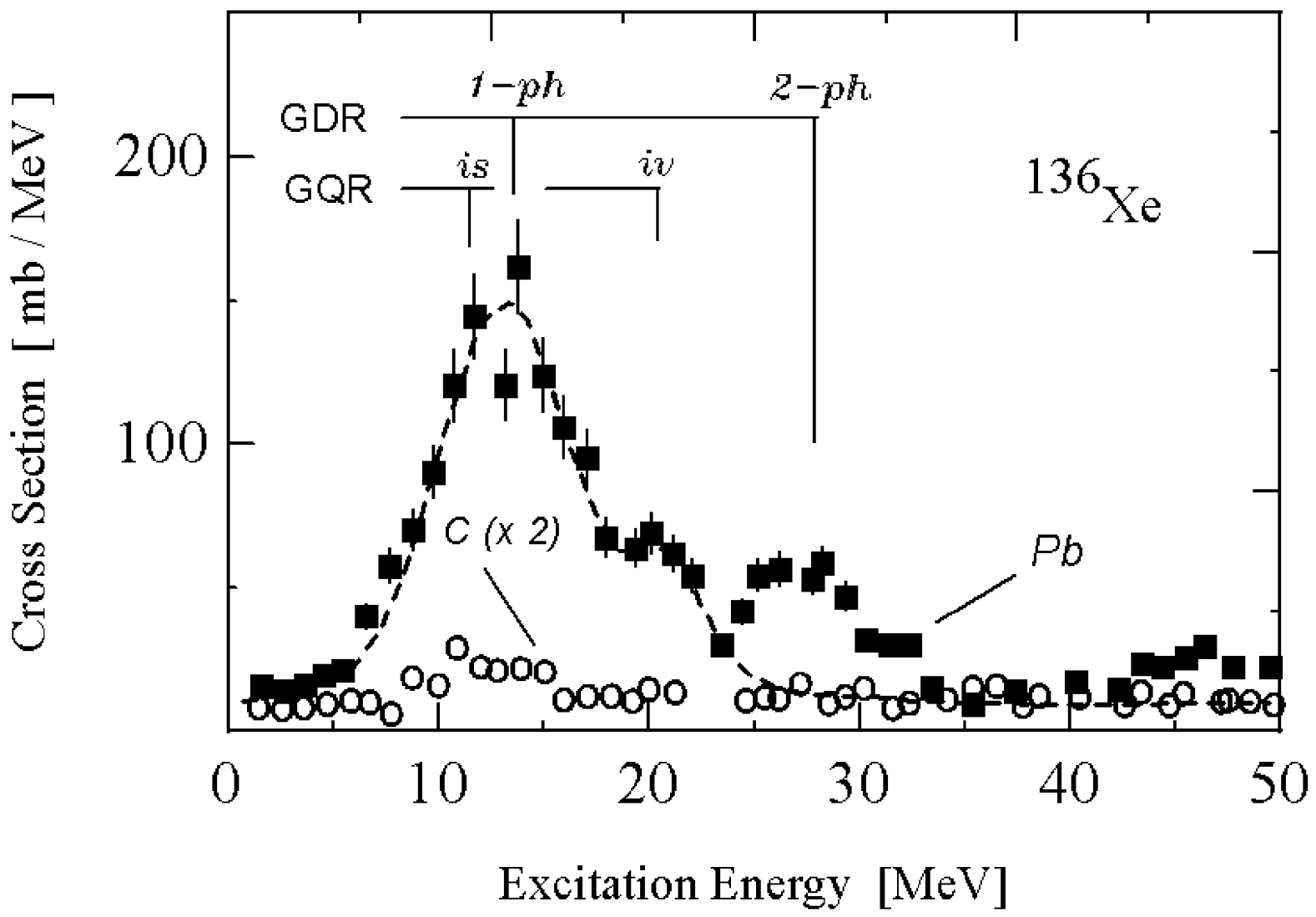

*Figure 3*

Assuming that one has $\sigma_\gamma^{\pi\lambda}(E)$ somehow (either from experiments, or from theory), a simple harmonic model based on the Axel-Brink hypotheses can be formulated to obtain the probability to access a multiphonon state of order $n$. In the harmonic oscillator model the inclusion of the coupling between all multiphonon states can be performed analytically [6]. One of the basic changes is that the excitation probabilities calculated to first-order, $P_{\pi\lambda}^{1st}(E, b)$, are modified to include the flux of probability to the other states. That is,

$$P_{\pi\lambda}(E,\ b) = P^{1st}_{\pi\lambda}(E,b)\ \exp\left\{-P^{1st}_{\pi\lambda}(b)\right\}\ , \tag{2}$$

where $P^{1st}_{\pi\lambda}(b)$ is the integral of over the excitation energy $E$. In general, the probability to reach a multiphonon state with the energy $E^{(n)}$ from the ground state, with energy $E^{(0)}$, is obtained by an integral over all intermediate energies

$$P^{(n)}_{\pi^*\lambda^*}(E^{(n)},b) = \frac{1}{n!}\ \exp\left\{-P^{1st}_{\pi\lambda}(b)\right\} \int dE^{(n-1)}\ dE^{(n-2)}\ ...\ dE^{(1)} \tag{3}$$
$$\times P^{1st}_{\pi\lambda}(E^{(n)}-E^{(n-1)},b)\ P^{1st}_{\pi\lambda}(E^{(n-1)}-E^{(n-2)},b)\ ...\ P^{1st}_{\pi\lambda}(E^{(1)}-E^{(0)},b)$$

The character and spin assignment of the multipolarity $\lambda^*$ depends on how the intermediate states couple with the electromagnetic transition operators. For example, in the case of the DGDR ($GDR^2$), assuming a $0^+$ ground state and excluding isospin impurities, the final state has either spin and parity $0^+$ or $2^+$, respectively.

A simpler reaction model than above can be obtained when assuming that all states can be approximated by a single isolated state. For example, we can assume that the photoabsorption cross sections in the range of the GDR is due to a single state with energy equal to the centroid energy of the GDR exhausting the whole excitation strength. Then the multiphonon states are equidistant, and eq. (3) becomes

$$P^{(n)}_{\pi^*\lambda^*}(b) = \frac{1}{n!}\left[P^{1st}_{\pi\lambda}(b)\right]^n\ \exp\left\{-P^{1st}_{\pi\lambda}(b)\right\}\ . \tag{4}$$

The above relation was used to calculate the cross sections for the excitation of the GDR, GDR$^2$, GDR$^3$, ISGQR and IVGQR in $^{136}$Xe, respectively, for collisions with Pb nuclei as a function of the bombarding energy, as shown in figure 2. Each resonance is considered to be a single state exhausting 100% of the respective sum rule. Also shown in the figure is the geometrical cross section (G.C.), $\sigma \sim \pi\left(A_P^{1/3}+A_T^{1/3}\right)^2\ fm^2$. The cross sections for the excitation of the GDR$^2$ is large, of order of hundreds of mb.

Much of the interest on looking for multiphonon resonances relies on the possibility for looking at exotic particle decay of these states. For example, in ref. [12] a hydrodynamical model was used to predict the proton and neutron dynamical densities in a multiphonon state of a nucleus. Large proton and neutron excesses at the surface are developed in a multiphonon state. Thus, the emission of exotic clusters from the decay of these states are a natural possibility. A more classical point of view is that the Lorentz contracted Coulomb field in a peripheral relativistic heavy ion collision acts as a hammer on the protons of the nuclei[7]. This (collective) motion of the protons seem only to be probed in relativistic Coulomb excitation. It is not well known how this classical view can be related to microscopic properties of the nuclei in a multiphonon state.

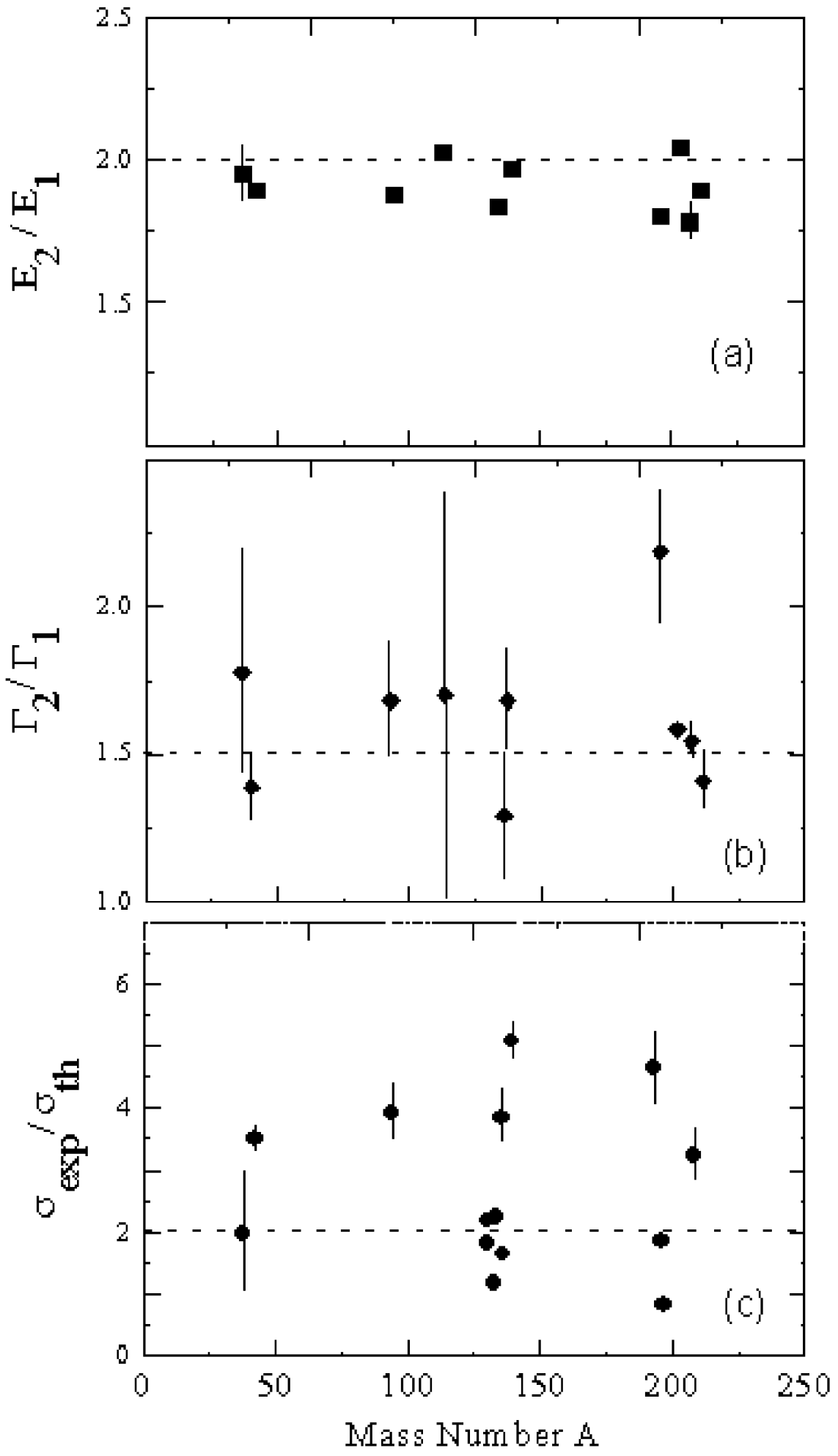

*Figure 4*

Although the perspectives for an experimental evidence of the DGDR via relativistic Coulomb excitation were good, on the basis of the large theoretical cross sections, it was first found in pion scattering at the Los Alamos Pion Facility[9]. In pion scattering off nuclei the DGDR can be described as a two-step mechanism induced by the pion-nucleus interaction. Using the Brink-Axel hypotheses, the cross sections for the excitation of the DGDR with pions were shown to be well within the experimental possibilities [9]. Only about 5 years later, the first Coulomb excitation experiments for the excitation of the DGDR were performed at the GSI facility in Darmstadt/Germany [10,11]. In figure 3 we show the result of one of these experiments, which looked for the neutron decay channels of giant resonances excited in relativistic projectiles. The excitation spectrum of relativistic $^{136}$Xe projectiles incident on Pb are compared with the spectrum obtained in C targets. A comparison of the two spectra immediately proofs that nuclear contribution to the excitation is very small. Another experiment [11] dealt with the photon decay of the double giant resonance. A clear bump in the spectra of coincident photon pairs was observed around the energy of two times the GDR centroid energy in $^{208}$Pb targets excited with relativistic $^{209}$Bi projectiles.

The advantages of relativistic Coulomb excitation of heavy ions over other probes (pions, nuclear excitation, etc.) was clearly demonstrated in several GSI experiments [10,11,13,14]. We now discuss many features of the double GDR that were obtained.

## 2 Energy, Width, and Strength of the Double GDR

A collection of the experimental data on the energy and width of the DGDR is shown in figure 4. The data points are from a compilation from pion, Coulomb excitation and nuclear excitation experiments [15].

The dashed lines are guide to the eyes. We see from figure 4(a) that the energy of the DGDR agrees reasonably with the expected harmonic prediction that the energy should be about twice the energy of the GDR, although small departures from this prediction are seen, especially in pion and nuclear excitation experiments. The width of the DGDR seems to agree with an average value of $\sqrt{2}$ times that of the GDR, although a factor 2 seems also to be possible, as we see from figure 4(b). Figure 4(c) shows the ratio between the experimentally determined cross sections and the calculated ones. Here is where the data appear to be more dispersed. The largest values of $\sigma_{\rm exp}/\sigma_{th}$ come from pion experiments, yielding up to a value of 5 for this quantity.

### *2.1 Width of the DGDR*

In a microscopic approach, the GDR is described by a coherent superposition of one-particle one-hole states. One of the many such states is pushed up by the residual interaction to the experimentally observed position of the GDR. This state carries practically all the E1 strength. This situation is simply realized in a model with a separable residual interaction. We write the GDR state as (one phonon with angular momentum 1M) $|1,\ 1M\rangle = A^{\dagger}_{1M}\,|0\rangle$ where $A^{\dagger}_{1M}$ is a proper superposition of particle-hole creation operators. Applying the quasi-boson approximation we can use the boson commutation relations and construct the multiphonon states (N-phonon states). A N-phonon state will be a coherent superposition of N-particle N-hole states. The width of the GDR is essentially due to the spreading width, i.e., to the coupling to more complicated quasibound configurations. The escape width plays only a minor role. We are not interested in a detailed microscopic description of these states here. We use a simple model for the strength function [16]. We couple a state $|a\rangle$ (i.e. a GDR state) by some mechanism to more complicated states $|\alpha\rangle$, for simplicity we assume a constant coupling matrix element $V_{a\alpha} = \langle a\,|V|\,\alpha\rangle = \langle \alpha\,|V|\,a\rangle = \mathrm{v}$. With an equal spacing of $D$ of the levels $|\alpha\rangle$ one obtains a width

$$\Gamma = 2\pi \frac{\mathrm{v}^2}{D}\ , \tag{5}$$

for the state $|\alpha\rangle$ . We assume the same mechanism to be responsible for the width of the N-phonon state: one of the N-independent phonons decays into the more complicated states $|\alpha\rangle$ while the other (N-1)-phonons remain spectators. We write the coupling interaction in terms of creation (destruction) operators $c^{\dagger}_{a}$ $(c_\alpha)$ of the complicated states $|\alpha\rangle$ as

$$V = \mathrm{v}\left(A^{\dagger}_{1M}\ c_\alpha + A_{1M}\ c^{\dagger}_{\alpha}\right)\ . \tag{6}$$

For the coupling matrix elements $\mathrm{v}_N$, which connects an N-phonon state $|N\rangle$ to the state $|N-1,\alpha\rangle$ (N-1 spectator phonons) one obtains

$$\mathrm{v}_N = \langle N-1,\alpha\,|V|\,N\rangle = \mathrm{v}\,\langle N-1\,|A_{1M}|\,N\rangle = \mathrm{v}\ .\ \sqrt{N}\ , \tag{7}$$

i.e., one obtains for the width $\Gamma_N$ of the N-phonon state

$$\Gamma = 2\pi N \frac{\mathrm{v}^2}{D}\ = N\Gamma\ , \tag{8}$$

where $\Gamma$ is given by eq. (5).

Thus, the factor N in (8) arises naturally from the bosonic character of the collective states. For the DGDR this would mean $\Gamma_2 = 2\Gamma_1$. The data points shown in figure 4(b) seem to favor a lower multiplicative factor.

We can also give a qualitative explanation for a smaller $\Gamma_2/\Gamma_1$ value. First we note that the value $\Gamma_2/\Gamma_1 = N$ can also be obtained from a folding procedure, as given by eq. (3). If the sequential excitations are described by Breit-Wigner (BW) functions $P_{\pi\lambda}(E)$ with the centroid $\mathcal{E}$ and the width $\Gamma$, the convolution (3) yields a BW shape with the centroid at $2\mathcal{E}$ for the DGDR and the total width of $2\Gamma_1$. However, if one uses gaussian functions (instead of BW ones) for the shape of one-phonon states, it is easy to show that one gets also a gaussian for the N-phonon shape, but with the width given by $\sqrt{N}\Gamma_1$. The later assumption seems inconsistent since the experimentalists use BW fits for the shape of giant resonances, in good agreement with the experimental data. But one can easily understand that the result $\sqrt{N}\Gamma_1$ is not restricted to a gaussian fit. For an arbitrary sequence of two excitation processes we have $\langle E \rangle = \langle E_1 + E_2 \rangle$ and $\langle E^2 \rangle = \left\langle (E_1 + E_2)^2 \right\rangle$; for uncorrelated steps it results in the addition in quadrature $(\Delta E)^2 = (\Delta E_1)^2 + (\Delta E_2)^2$ . Identifying these fluctuations with the widths up to a common factor, we get for identical phonons $\Gamma_2 = \sqrt{N}\Gamma_1$. The same conclusion will be valid for any distribution function which, as the gaussian one has a finite second moment, contrary to the BW or lorentzian ones with second moment diverging. We may conclude that, in physical terms, the difference between $\Gamma_2/\Gamma_1 = 2$ and $\Gamma_2/\Gamma_1 = \sqrt{2}$ is due to the different treatment of the wings of the distribution functions which reflect small admixtures of far remote states.

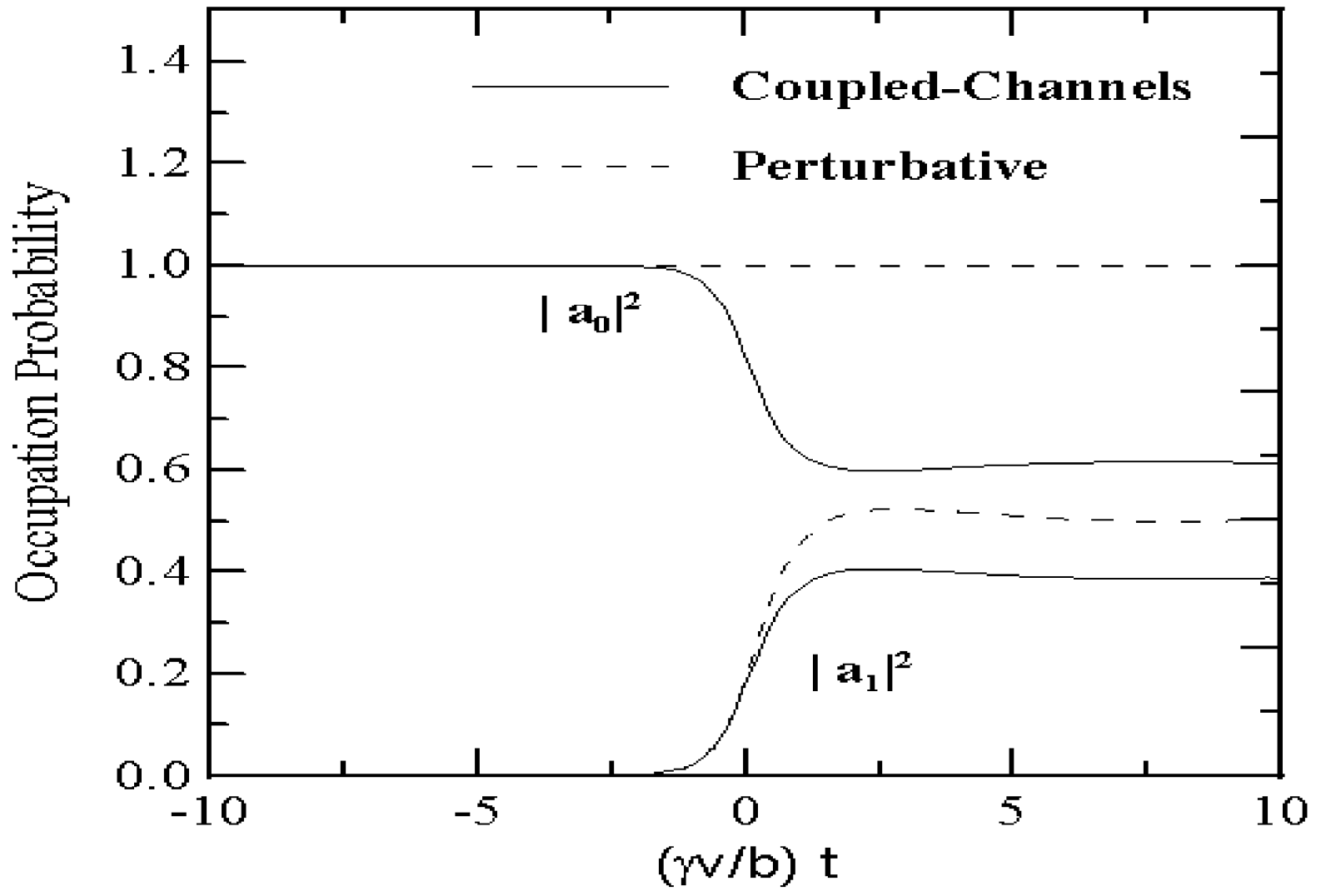

*Fig. 5*

*2.2 Strength of the DGDR*

Microscopically, the harmonic picture is accomplished within the RPA approximation. The excited states of the nucleus are described as superpositions of particle-hole configurations with respect to the ground state. The multi-phonon resonances are built by products of the $1^-$ resonance states, yielding $0^+$ and $2^+$ double phonon states. The interaction with the projectile is described in terms of a linear combination of particle-hole operators weighted by the time-dependent field for a given multipolarity of the interaction. Since the time dependent Coulomb field of a nucleus does not carry monopole multipolarity, the DGDR states can be reached via two-step E1 transitions and direct E2 transitions (for a $0^+$ ground state). As we see from figure 4(c), early calculations failed to explain the experimental data. Due to the simpler excitation mechanism we restrict ourselves to the Coulomb excitation cross sections. Then, there seems to be two possible reasons for $\sigma_{\rm exp}/\sigma_{th} \neq 1$; (a) either the Coulomb excitation mechanism is not well described, or (b) the response of the nucleus to two-phonon excitations is not well known.

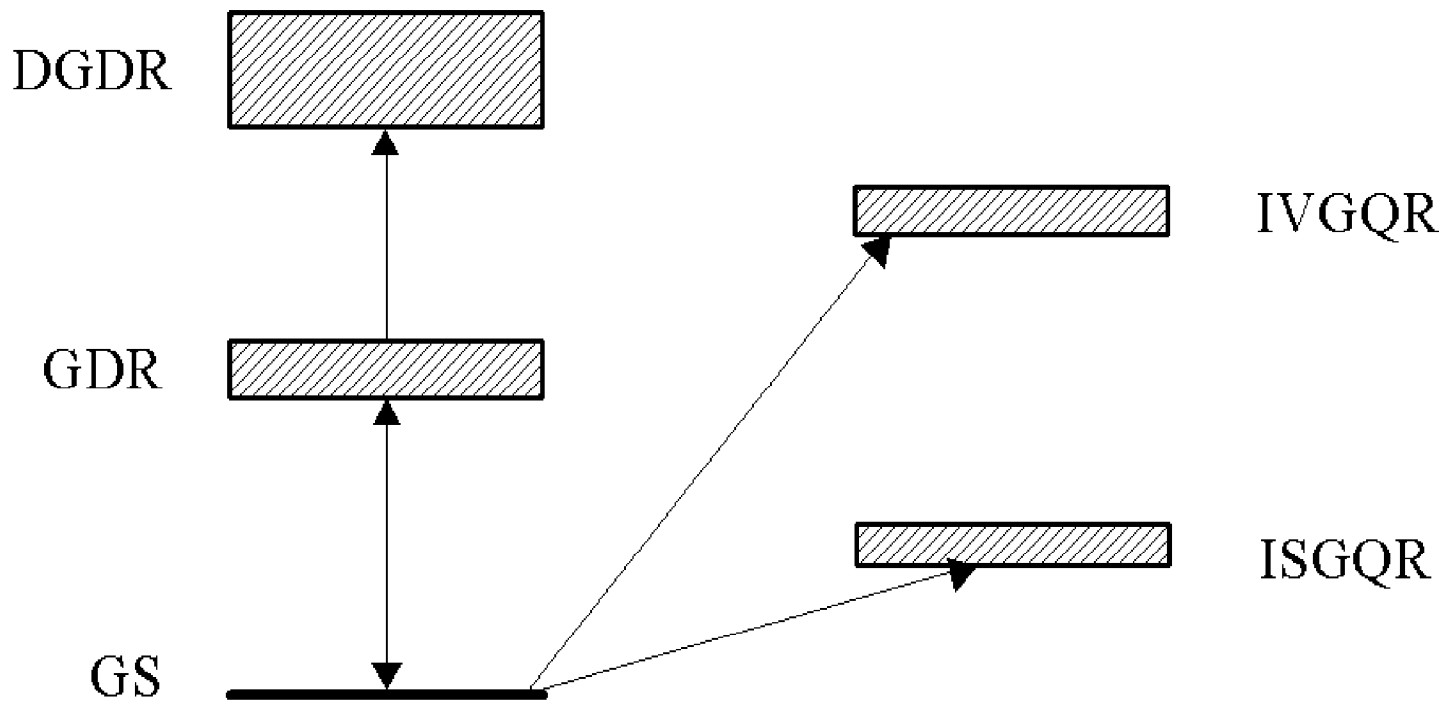

*Figure 6*

Many authors studied the effects of the excitation mechanism for the excitation of the DGDR. In ref. [17] the cross sections were calculated using the second order perturbation theory. It was found that the theoretical values were smaller than the experimental ones by about a factor of 1.3 - 2. However, it was suggested[18] that second order perturbation theory is not adequate for relativistic Coulomb excitation of GR's with heavy ions and that it is necessary to perform a coupled channels calculation. We see the this more clearly from figure 5, taken from ref. [19], where a coupled-channels study of multiphonon excitation by the nuclear and Coulomb excitations in relativistic heavy ion collisions was performed. The figure shows the probability amplitude to excite the GDR in $^{208}$Pb, $|a_1|^2$, and the occupation probability of the ground state, $|a_0|^2$, for a grazing collision of $^{208}$Pb+$^{208}$Pb at 640 MeV/nucleon. The dashed lines are the predictions of the first-order perturbation theory. We see that the asymptotic excitation probability of the GDR is quite large ($\sim$ 40%). In first-order perturbation the occupation probability of the ground-state is kept constant, equal to unity. Obviously, one greatly violates the unitarity condition in this case. A more appropriate coupled-channels calculation (solid lines) shows that the ground-state occupation probability has to decrease to meet the unitarity requirements, while the excitation probability of the GDR is also a bit reduced for the same reason.

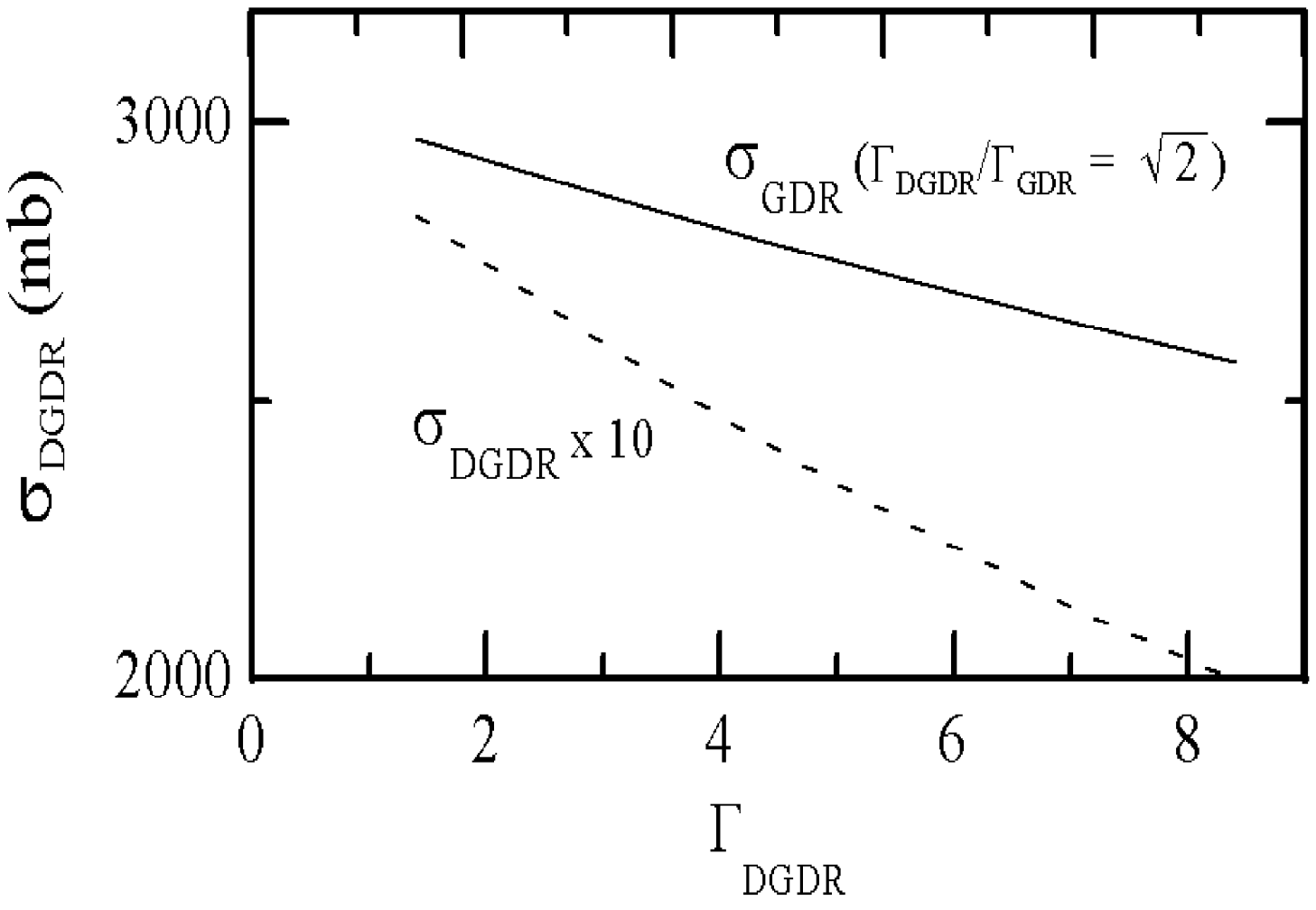

*Figure 7*

In ref. [19] it was shown that a good coupled-channels calculation does not need to account for the exact coupling equations in all channels. The strongest coupling, responsible for the effect observed in figure 5 is the coupling between the ground-state and the GDR states. This has to be treated exactly within a coupled-channels calculation. The coupling between the GDR and the other states (including the DGDR, IVGQR, ISGQR, etc.) can be treated perturbatively, as shown schematically in figure 6. This amounts in a great simplification of the calculation. In fact, it allows to include the width of the GDR in the coupled-channels calculation straightforwardly using a Breit-Wigner strength function for the GDR. In terms of the auxiliary amplitudes $A_\mu(t)$, given by the relation $a_0(t) = 1 + \sum_\mu A_\mu(t)$, with $\mu = -1, 0, 1$, the coupled-channels equations are given by

$$\ddot{A}_\mu(t) - \left[\frac{\dot{V}_\mu^{(01)}(t)}{V_\mu^{(01)}(t)} - \frac{i}{\hbar}\Big(E_1 - i\frac{\Gamma_1}{2}\Big)\right] \dot{A}_\mu(t) + \mathcal{S}_1 \frac{|V_\mu^{(01)}(t)|^2}{\hbar^2} \Big[1 + \sum_{\mu'} A_{\mu'}(t)\Big] . \tag{9}$$

where the factor $\mathcal{S}_1$ is $\mathcal{S}_1 = 1$ for BW-shape and $\mathcal{S}_1 = 1 - i\Gamma_1/2E_1$ for Lorentzian-shape, and $V_\mu^{(01)}(t)$ is the time-dependent Coulomb interaction between the ground-state and the magnetic component $\mu$ of the GDR. With initial conditions $A_\mu(t = -\infty) = 0$, the solution of the above coupled-channels equation can be used to calculate the excitation probability of the GDR state with energy $E_1 + \epsilon$, given by

$$i\hbar\ \dot{a}^{(1)}_{\epsilon,1\mu}(t) = \left[(\alpha^{(1)}(\epsilon)\ V_\mu^{(01)}(t)\right]^*\ \exp\left\{i(E_1+\epsilon)t/\hbar\right\}\ a_0(t)\ . \qquad (10)$$

where $\alpha^{(1)}(\epsilon)$ is a BW or Lorentzian shape function, with centroid at $E_1$. After the calculation of the occupation probability of the ground state, $|a_0\,(\infty)\,|^2$, and the excitation probability of the GDR, $\sum_\mu \int |a_{\epsilon,1\mu}\,(\infty)\,|^2 d\epsilon$ , it is straightforward to calculate the excitation probabilities of the DGDR and of other giant resonances using perturbation theory, as shown schematically in fig. 6. Excitation cross sections are obtained by integration over impact parameters.

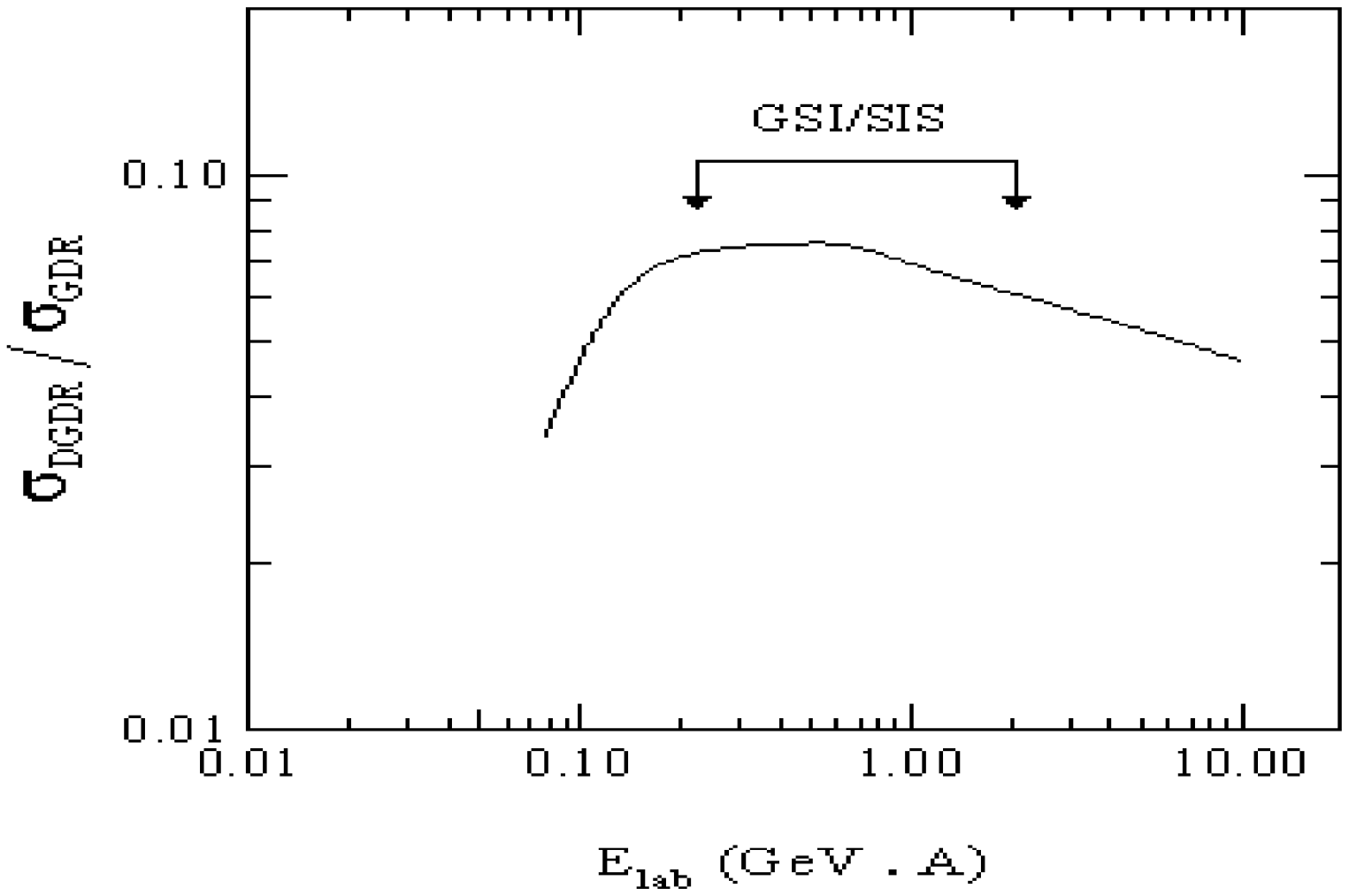

*Figure 8*

The results of ref. [19] showed appreciable dependence of the excitation cross sections of the GDR[2] on the width of both the GDR and the DGDR as can be seen in fig. 7 for $^{208}$Pb+$^{208}$Pb at 640 MeV/nucleon. The solid curve shows the GDR cross section as a function of the width of the DGDR, keeping the ratio $\Gamma_{DGDR}/\Gamma_{GDR} = \sqrt{2}$. The dashed curve is obtained by fixing the value of $\Gamma_{GDR} = 4$ MeV and varying the value of $\Gamma_{DGDR}$ . The cross sections decrease with energy since an increase of the width enhances the doorway amplitude to higher energies where Coulomb excitation is weaker. Based on this figure, we may conclude that the data seem to favor a value of $\Gamma_{DGDR}/\Gamma_{GDR} \simeq \sqrt{2}$. Figure 8 shows the ratio $\sigma_{DGDR}/\sigma_{GDR}$ as a function of the bombarding energy. We observe that the most favorable energies for the measurement of the DGDR corresponds to the SIS energies at the GSI-Darmstadt facility.

### *2.3 Anharmonicities*

Another possible effect arises from a shift of the energy centroid of the DGDR due to anharmonic effects [20]. In ref. [19] one obtained $\sigma_{DGDR} = 620$ mb, 299 mb, and 199 mb for the centroid energies of $E_{DGDR} = 20$ MeV, 24 MeV and 27 MeV, respectively. This shows that anharmonic effects can play a big role in the Coulomb excitation cross sections of the DGDR, depending on the size of the shift of $E_{DGDR}$. However, in ref. [17] the source for anharmonic effects were discussed and it was suggested that it should be very small, i.e., $\Delta^{(2)}E = E_{DGDR} - 2E_{GDR} \simeq 0$.

The anharmonic behavior of the giant resonances as a possibility to explain the increase of the Coulomb excitation cross sections has been studied by several authors [20,21] (see also ref. [22], and references therein). It was found that the effect is indeed negligible and it could be estimated [22] as $\Delta^{(2)}E < E_{GDR}/(50.A) \sim A^{-4/3}$ MeV.

### *2.4 Other routes to the DGDR*

From the discussion above we see that the magnitude of the Coulomb excitation cross sections of the DGDR can be affected due to uncertainties in: (a) strength, (b) width, (c) energies, or (d) reaction mechanism. Case (a) and (c) are the basis of the Brink-Axel hyphotesis and we have seen that a modification of their values would only be considered seriously if anharmonic effects were large, which seems not to be the case. Case (b) is an open question. Microscopic calculations [21] have shown that, taking into account the Landau damping, the collective state splits into a set of different $1_i^-$ states distributed over an energy interval, where $i$ stands for the order number of each state. A further fragmentation of the $1_i^-$ states into thousands of closed packed states, is obtained by the coupling of one-phonon and two-phonon states. This leads to a good estimate of the spreading width of the GDR. However, the DGDR states were obtained by a folding procedure:

$$|[1_i^- \otimes 1_{i'}^-]_{J^\pi=0^+,1^+,2^+} >_M = \sum_{m,m'} (1m1m'|JM)|1_i^- >_m |1_{i'}^- >_{m'}, \qquad (11)$$

The width of the DGDR is thus fixed from the width of the GDR. Critics should be remembered that even the spreading width of the giant dipole resonance is not very well described theoretically. It is therefore impossible to make any quantitative prediction for the width of the DGDR, other than saying that $\sqrt{2} \leq \Gamma_{DGDR}/\Gamma_{GDR} \leq 2$.

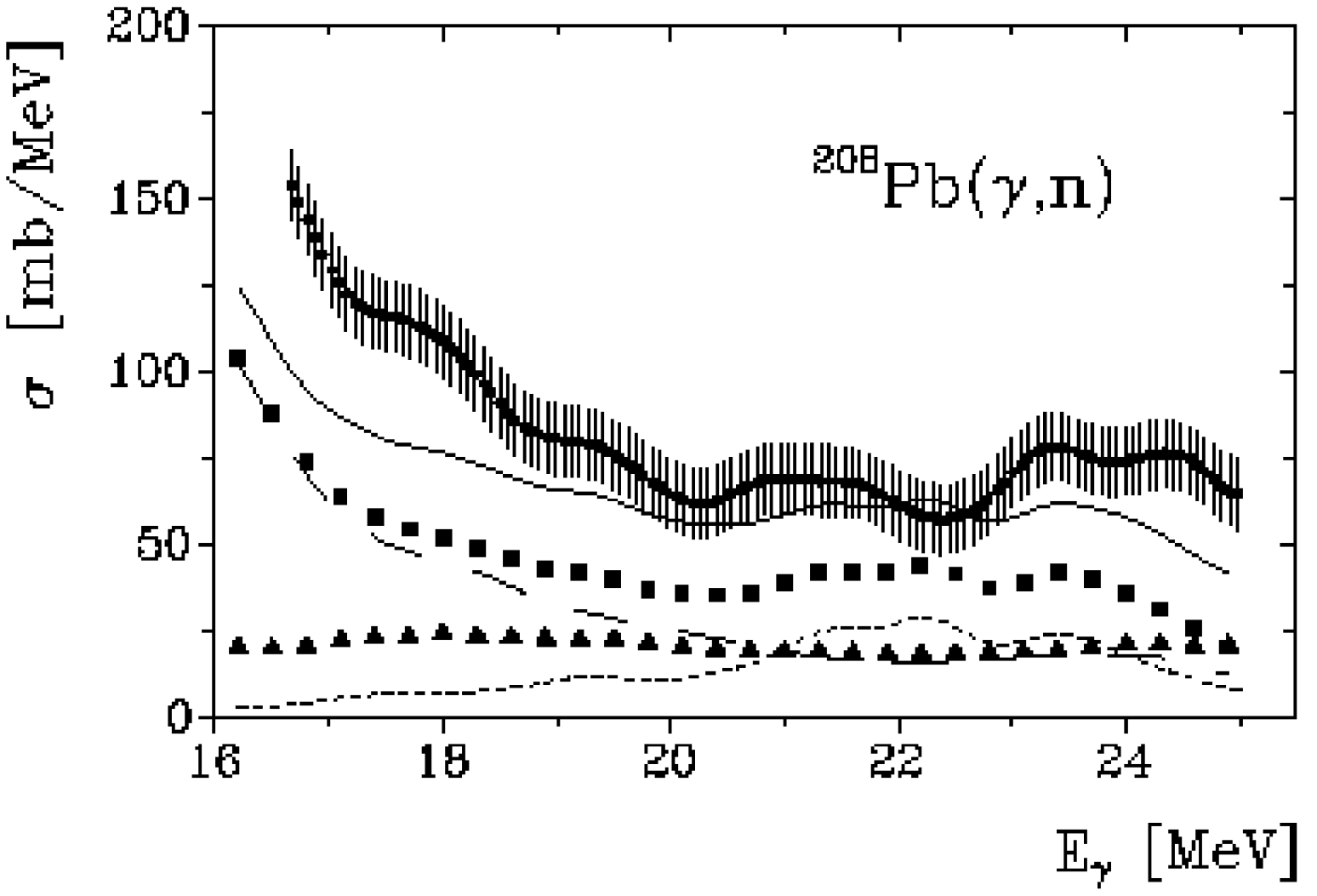

*Fig. 9*

We return to the discussion of the reaction mechanism, and how it could affect the magnitude of the cross sections. It looks obvious from figure 3 that the nuclear excitation of giant resonances is very small in magnitude compared to Coulomb excitation in collisions with heavy ions at relativistic energies. However, the nuclear-Coulomb interference could also be relevant and would not be in complete disagreement with the experimental findings. But, in ref. [19] it was shown that this is a small effect indeed.

We have seen that coupled-channels effects are very important and should always be considered in the analysis of the experimental results. In ref. [23] the contribution of non-natural parity $1^+$ two-phonon states were investigated in a coupled-channels calculation. The diagonal components $[1_i^- \otimes 1_i^-]_{1^+}$ are forbidden by symmetry properties but nondiagonal ones $[1_i^- \otimes 1_{i'}^-]_{1^+}$, a priori, may be excited in two-step process bringing some "extra strength" in the DGDR region. Consequently, the role of these nondiagonal components depends on how strong is the Landau damping.

The coupled-channels calculation found that the contribution of the $1^+$ states to the total cross section is small. The reason for this is better explained in second-order perturbation theory. For any route to a final magnetic substate $M$, the second-order amplitude will be proportional to $(001\mu|1\mu)V_{E1\mu,0\to1^-}\times(1\mu1\mu'|1M)\ V_{E1\mu',1^-\to1^+}+(\mu\longleftrightarrow\mu')$, where $V_{E1\mu,i\to f}$ is the $\mu$-component of the interaction potential (for a spin-zero ground state, $\mu$ is also the angular momentum projection of the intermediate state). Assuming that the phases and the products of the reduced matrix elements for the two sequential excitations are equal, we get $V_{E1\mu,0\to1^-}\times V_{E1\mu',1^-\to1^+}=V_{E1\mu',0\to1^-}\times V_{E1\mu,1^-\to1^+}$. Thus, under these circumstances, and since $(001\mu|1\mu)\equiv1$, we get an identically zero result for the excitation amplitude of the $1^+$ DGDR state as a consequence of $(1\mu1\mu'|1M)=-\ (1\mu'1\mu|1M)$. A coupled-channels calculation cannot change this result appreciably.

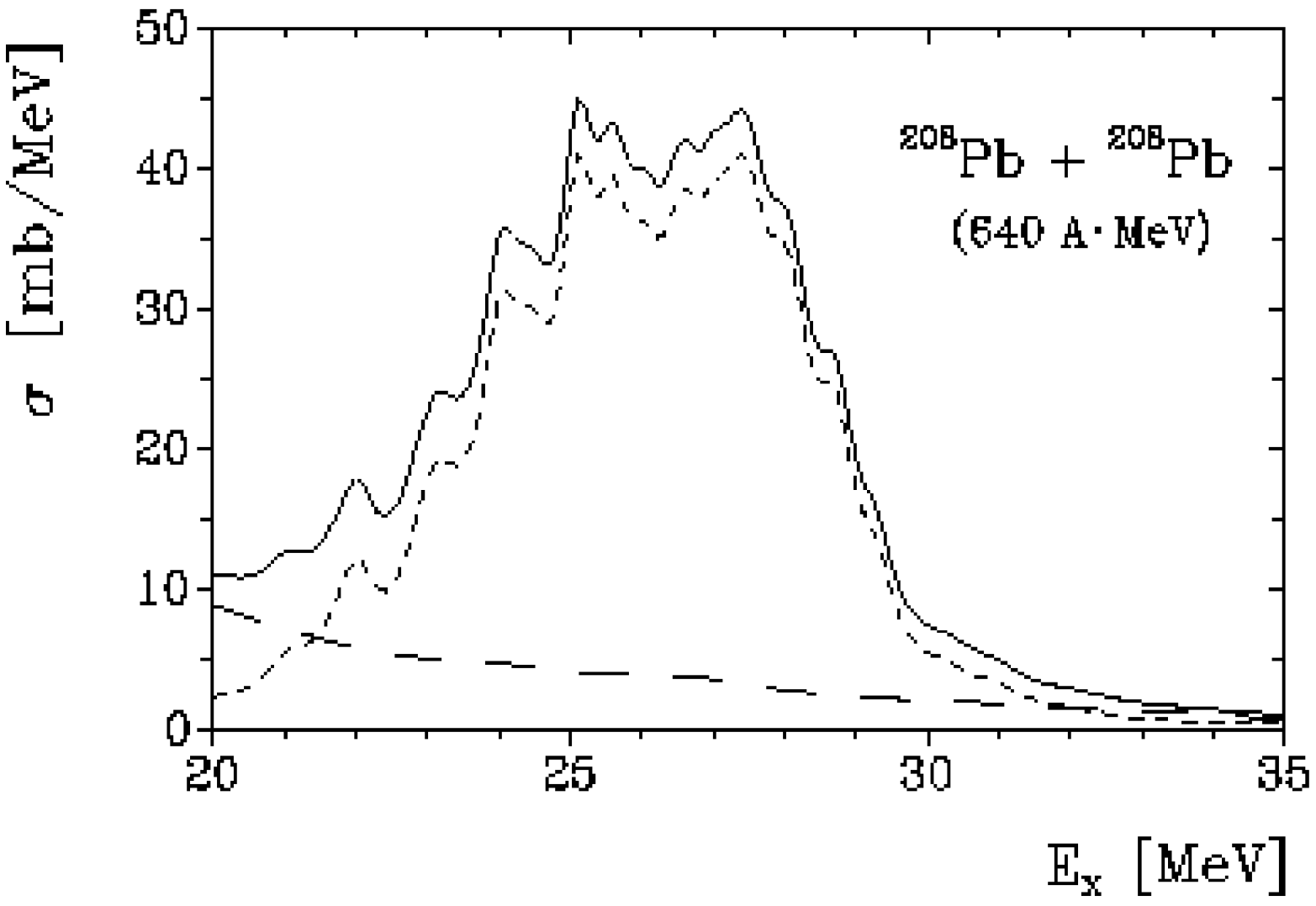

*Fig. 10*

We note that multiphonon states can be obtained by coupling all kinds of phonons. Each configuration $[\lambda_1^{\pi_1}\otimes\lambda_2^{\pi_2}]$ can be obtained theoretically from a sum over several two-phonon states made of phonons with a given spin and parity $\lambda_1^{\pi_1}$, $\lambda_2^{\pi_2}$, and different RPA root numbers $i_1$, $i_2$ of its constituents. The cross sections can be obtained accordingly:

$$\sigma([\lambda_1^{\pi_1}\otimes\lambda_2^{\pi_2}])=\sum_{i_1,i_2}\sigma([\lambda_1^{\pi_1}(i_1)\otimes\lambda_2^{\pi_2}(i_2)])\tag{12}$$

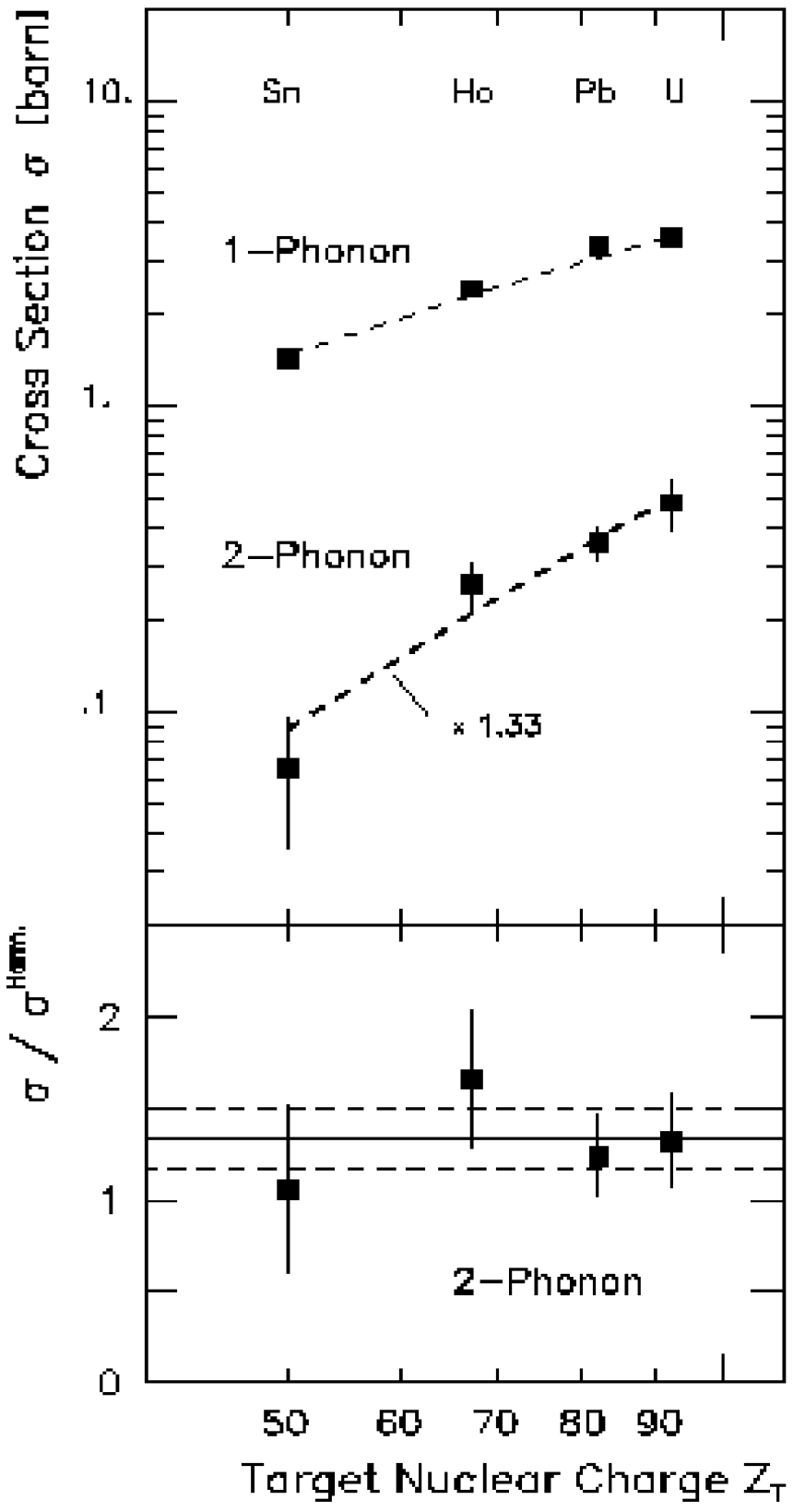

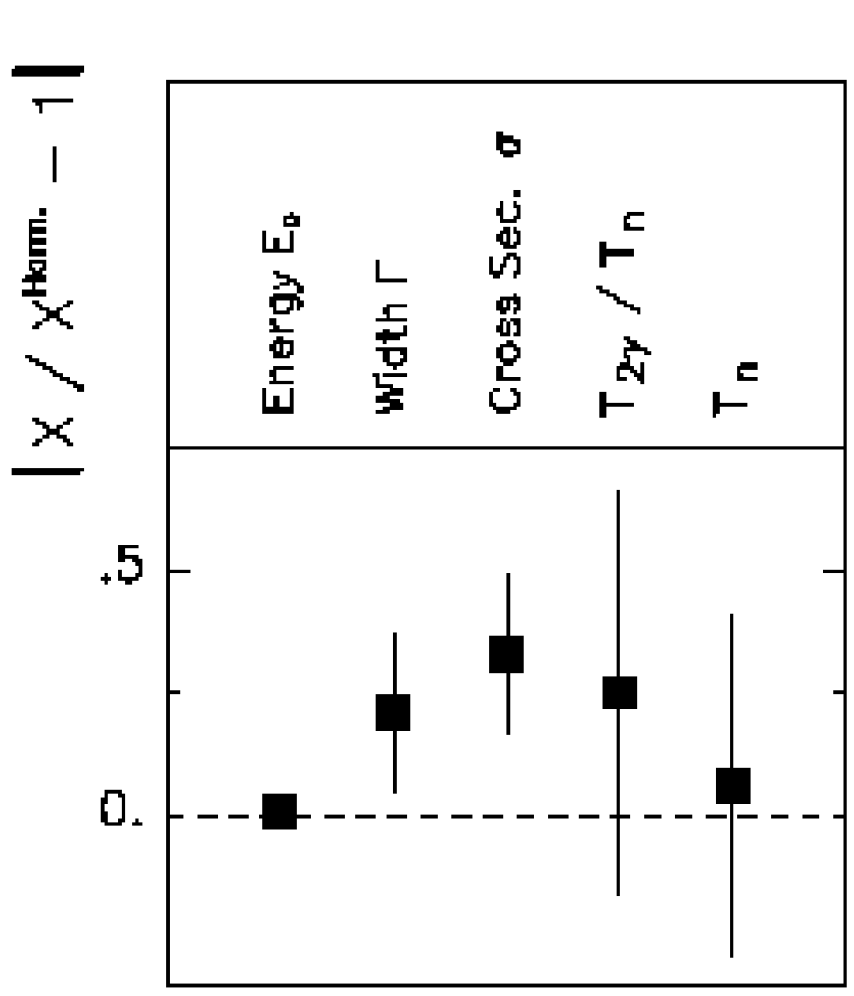

*Figure 11(a)*                    *Figure 11(b)*

As an example, in ref. [24] the total number of two-phonon $1^-$ states generated in this way was about $10^5$. The absolute value of the photoexcitation of any two-phonon state under consideration is negligibly small but altogether they produce a sizable cross section. The $1^-$ two-phonon states obtained in ref. [24] were used to calculate their contribution to the $(\gamma, n)$ cross section in $^{208}$Pb, via direct E1 excitations. This is shown in figure 9. Experimental data (dots with experimental errors) are from ref. [25]. The long-dashed curve is the high energy tail of the GDR, the short-dashed curve is the IVGQR and the curve with squares is their sum. The contribution of two-phonon $1^-$ states is plotted by a curve with triangles. The solid curve is the total calculated cross section. Thus, already at the level of photonuclear data the contribution of two-phonon $1^-$ states is of relevance. Here they are not reached via two-step processes, but in direct excitations. Since the energy region of these states overlap with that of the DGDR, in Coulomb excitation experiments they should also contribute appreciably. In fact, it was shown recently [24] that their contribution to the total cross section for $^{208}$Pb + $^{208}$Pb (640A MeV) in the DGDR region is of order of 15%. In figure 10 the contribution for the excitation of two-phonon $1^-$ states (long-dashed curve) in first order perturbation theory, and for two-phonon $0^+$ and $2^+$ DGDR states in second order (short-dashed curve). The total cross section (for $^{208}$Pb (640 A·MeV) +$^{208}$Pb) is shown by the solid curve.

### *2.5 Present situation and perspectives*

The experimental situation on the excitation of double phonon states has improved considerably, mainly due to advances on the data acquisition with the Coulomb excitation technique. Figure 11(a) shows the comparison between the experimental cross section and the simple harmonic model (eqs. 2 and 3) for the excitation of the GDR and the DGDR in $^{208}$Pb (640A MeV) and several targets [14]. One sees that the harmonic model reproduces quite well the cross section for the GDR, but it misses the magnitude of the cross section by 30%, as can be better seen in the lower part of the figure where the ratio between the experimental data and the calculation is shown. In figure 11(b) the present situation on the knowledge of the energy, width, excitation cross section, branching ratio for gamma to neutron emission, and the neutron emission width, respectively, is shown in comparison with calculations based on the simple harmonic picture. We see that the theory-experiment agreement is much better than those presented in figure 4.

As we have seen in this short review there are several effects which compete in the excitation of double giant resonances in relativistic Coulomb excitation. These effects were discovered in part by the motivation to explain discrepancies between the harmonic picture of the giant resonances and the recent experimental data. We cannot say at the moment by how much we have progressed towards a better understanding of these nuclear structures. But, we can surely say that the field is just at its infancy and important experimental and theoretical progress will be underway in next future.

## 3 Acknowledgments

I am grateful to Vladimir Ponomarev for useful comments and suggestions. This work was partially supported by the FUJB/UFRJ, and by the MCT/FINEP/CNPq(PRONEX) (contract 41.96.0886.00).